# Efficient acceleration of a dense plasma projectile to hyper velocities in the laser-induced cavity pressure acceleration scheme

J.Badziak[1x], E.Krousky[2,3], J.Marczak[4], P.Parys[1], T.Pisarczyk[1], M.Rosiński[1], A.Sarzynski[4],
T.Chodukowski[1], J.Dostal[2,3], R.Dudzak[2], Z.Kalinowska[1], M.Kucharik[5], R.Liska[5], M.Pfeifer[2,3],
J.Ullschmied[2,3], A.Zaras-Szydłowska[1]

[1]*Institute of Plasma Physics and Laser Microfusion, 01-497, Warsaw, Poland*;
[2]*Institute of Plasma Physics ASCR, 182 00, Prague, Czech Republic;*
[3]*Institute of Physic ASCR s, 182 21, Prague, Czech Republic;*
[4]*Military University of Technology in Warsaw, 01-476, Warsaw,Poland*
[5]*Czech Technical University, FNSPE, 166 36, Prague, Czech Republic,*

The experimental study of the plasma projectile acceleration in the laser-induced cavity pressure acceleration (LICPA) scheme is reported. In the experiment performed at the kilojoule PALS laser facility, the parameters of the projectile were measured using interferometry, a streak camera and ion diagnostics, and the measurements were supported by two-dimensional hydrodynamic simulations. It is shown that in the LICPA accelerator with a 200-J laser driver, a 4-μg gold plasma projectile is accelerated to the velocity of 140 km/s with the energetic acceleration efficiency of 15 – 19 % which is at least several times higher than those achieved with the commonly used ablative acceleration and the highest among the ones measured so far for any projectiles accelerated to the velocities ≥ 100 km/s. This achievement opens the possibility of creation and investigation of high-energy-density matter states with the use of moderate-energy lasers and may also have an impact on the development of the impact ignition approach to inertial confinement fusion.

## 1. Introduction

Acceleration of solid or dense plasma projectiles to hyper velocities is a topic of high relevance for contemporary research in high energy density physics (HEDP) [1,2] and inertial confinement fusion (ICF) [3,4]. Moreover, it finds applications in space research [5,6] and materials science, particularly in the studies of materials subject to high mechanical loads [2,7]. To achieve the projectile parameters useful for these research, various accelerating devices and acceleration schemes have been proposed and investigated. These include Van de Graaf accelerators [5], light-gas guns [2,7], pulsed-power machines [8] and laser-driven accelerators [3,4,9-15]. Among these, the laser-driven accelerators generate projectiles (usually dense plasma projectiles) of highest velocities up to ~ 100 – 1000 km/s [4,12] and have the potential to achieve even sub-relativistic projectile velocities [16,17]. The laser-based scheme most commonly used for the acceleration of plasma projectiles is the so-called ablative acceleration (AA) [3]. In this scheme, the surface of a solid target is irradiated by the laser beam or the laser-produced soft X-ray radiation, which leads to the creation of hot plasma expanding backward, thus accelerating the remaining denser part of the target (the projectile) in the forward direction via the "rocket effect" [3]. In particular, the AA scheme is used to drive and compress the hydrogen fuel in fusion targets [3,18] or to accelerate a projectile to ignite the compressed fuel by the impact [4,19]. Unfortunately, this scheme has low energetic efficiency of the acceleration $\eta_{acc} = E_p/E_L$ ($E_p$ is the kinetic energy of the projectile and $E_L$ is the laser energy), which in practice attains values in the range of 0.5 – 3 % [4,10-15,19]. As a result, to accelerate to high velocities (≥ 100 km/s) a projectile of relatively big mass ($m_p$ ~ 1 – 100 μg), a large kJ or multi-kJ laser driver is required [4, 10-12,15,18,19].

[x]Electronic address: jan.badziak@ifpilm.pl

Recently, Badziak et al. have proposed [20,21] a new scheme of acceleration of dense plasma capable of accelerating a plasma projectile with the energetic efficiency higher than that achieved in the AA scheme. In this scheme, referred to as the laser-induced cavity pressure acceleration (LICPA), a projectile (e.g. a small disc) placed in a cavity is irradiated by a laser beam introduced into the cavity through a hole and then accelerated in a guiding channel by the pressure created in the cavity by the laser-produced hot plasma. In this paper, using three different diagnostic methods we demonstrate that in the LICPA accelerator with a sub-ns 200-J laser driver, a 4-μg gold plasma projectile can be accelerated to the hyper velocity ~ 140 km/s with the energetic efficiency of 15 – 19 % which is the highest among the ones measured so far for any projectiles accelerated to the velocities ≥ 100 km/s.

## 2. Results and discussion

The experiment was performed at the kilojoule PALS laser facility [22] in Prague. Since one of the aim of this experiment was to compare the efficiency of acceleration in the LICPA scheme with that in the AA scheme whose efficiency is higher for short-wavelength laser drivers, we used the 3ω (λ = 0.438 μm), 0.3-ns PALS beam to drive the projectile in both schemes. In the LICPA case, a 3 μm thick gold disc covered with a 5 μm CH ablator was irradiated by the beam of energy about 200 J inside the cavity of the cylindrical LICPA accelerator characterized by the following parameters: the cavity length $L_c$ = 0.4 mm, the channel length $L_{Ch}$ = 2 mm, the channel diameter $D_{Ch}$ = 0.3 mm and the cavity hole diameter $D_h$ = 0.15 mm (Fig.1). The laser beam diameter on the CH/Au target surface was equal to ~ 300 μm and the laser intensity was ~ $10^{15}$ W/cm$^2$. For measurements of parameters of the plasma projectile three different diagnostics were applied: the three-frame optical interferometry [23] using the 2ω PALS laser beam, the time-of-flight (TOF) measurements employing the set of nine ion charge collectors [24,25] and the streak camera Hamamatsu C7700 with the time resolution ~ 10 ps. In the AA case, the 5-μm CH foil with the attached 3-μm Au disc was mounted on the entrance of a large cylindrical channel with the length $L_{Ch}$ = 2 mm and the diameter of 2 mm. In both schemes the laser beam diameter and intensity on the CH/Au target surface were approximately the same. For measurements of projectile parameters in the AA scheme we applied the interferometry and the TOF ion diagnostic (using the streak camera was impossible since the gold projectile was proceeded by an intense flux of CH plasma driven by the laser beam "wings").

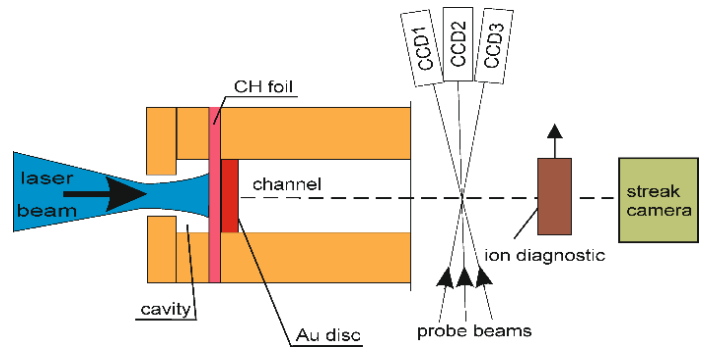

Fig. 1. A simplified scheme of the experimental set-up for measurements of parameters of the plasma projectile accelerated in the LICPA scheme. The ion diagnostics comprises nine ion charge collectors located approximately symmetrically relative to the accelerator axis (see the text for details).

The interferometry enabled us to measure the electron density distribution of plasma leaving the LICPA accelerator channel in three time points spaced by 3 ns. To calculate the electron density distributions for each the time point, phase shifts were extracted from the recorded interferograms using the maximum-of-the-fringe method [23]. Fig. 2 presents sample results of interferometric measurements and particularly the electron density distributions for the plasma flowing out of the accelerator channel recorded 20 ns after the CH/Au target irradiation by the laser pulse. It can be seen a collimated plasma flux with a high-density part (the opacity zone and neighborhood) of the transverse size equal to $D_{Ch}$. Based on the measurements like those presented in Fig. 2, we can estimate the effective (averaged in space) velocity of the plasma projectile leaving the accelerator channel from the expression $v_p \approx \beta \langle v_p \rangle$ , where $\langle v_p \rangle = L_{acc}/t_{acc}$ is the average velocity of the projectile on the acceleration length $L_{acc}$, $t_{acc}$ is the time of acceleration and $\beta$ is a correction factor whose value depends on how the acceleration $a$ changes in time (or along the acceleration path z). In case when the pressure driving the projectile is constant and, as a result, $a$ = const., the factor $\beta$ = 2. In the LICPA scheme, the pressure changes in time (and along z) during the acceleration process in the way that could not be determined from measurements. That is why to determine this factor we used numerical simulations employing the two-dimensional (2D) hydrodynamic PALE code [26,27]. Based on the simulations performed for the parameters of the CH/Au target, the accelerator and the laser pulse such as those in the experiment (for more details see [28]) we found that for $E_L$ = 200 J the factor $\beta \approx 1.36 \pm 0.02$. These simulations also indicated that the moment the projectile is leaving the accelerator channel about 90 % of the mass and kinetic energy of the projectile is stored inside a small region of the size $\Delta z_p$ < 0.1 mm (Fig. 5). Thus, we can assume (with an uncertainty $\Delta t \approx \Delta z_p/v_p$ < 1ns) that the moment when the high-density plasma, which is represented in the electron density distribution in Fig.2 by the opacity zone, appears on the channel outlet corresponds to the moment when the projectile is leaving the channel. Then, the acceleration length and the acceleration time can be written in the following form: $L_{acc} = L_{Ch} + \Delta z_{op}$, $t_{acc} = \Delta t_r - \Delta t_p$, where $\Delta z_{op}$ is the opacity zone length, $\Delta t_r$ is the time delay between the main laser pulse and the probe beam used for creation of the interferogram and $\Delta t_p$ is the time period of pressure build-up in the accelerator cavity. In case of our experiment, the value $\Delta t_p$ is small compared to $\Delta t_r$ and can be determined only roughly without significant influence on $t_{acc}$. For our experimental conditions we have $\Delta t_p \sim 2L_c/v_{cs} \approx 1.6$ ns, where $v_{cs} \approx 5 \times 10^7$ cm/s is the sound velocity in the cavity plasma (determined from the PALE simulations) and $L_c$ = 0.4 mm. Thus, for the case presented in Fig. 2 we obtain $\langle v_p \rangle \approx 114$ km/s, $v_p \approx 155$ km/s.

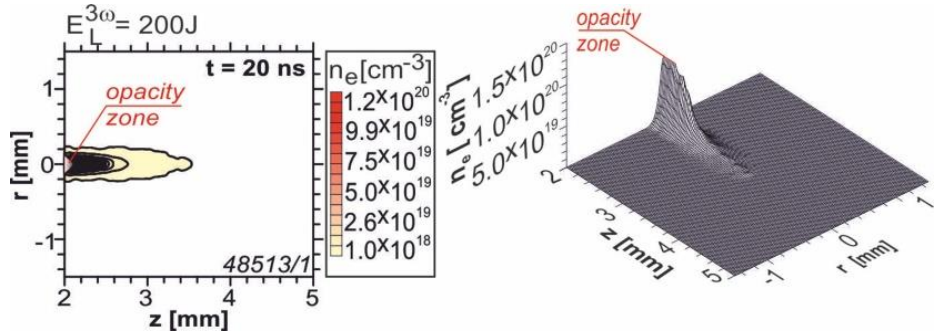

Fig. 2. The electron isodensitogram (left) and the space profile of electron distribution (right) for the plasma flowing out of the LICPA accelerator channel recorded 20 ns after the CH/AU target irradiation.

To determine the projectile velocity with the use of the streak camera, the method usually referred to as shock breakout chronometry (e.g. [10,15]) was applied. For this measurements the LICPA accelerator channel outlet was closed by a 20 um thick Al foil and the set of ion collectors was removed from the path between the accelerator and the camera (Fig. 1). We observed the light emission from the rear side of the foil which was imaged by an optical system on the entrance slit of the camera. The beginning of the emission occurred when the shock wave generated by the impact of the Au projectile into foil reached the foil rear surface. This shock breakout time ($t_{sb}$) was measured with respect to the so-called fiducial: a portion of the main laser pulse with a known delay with respect to the time ($t_L$) when the main pulse reaches the target in the accelerator cavity. A sample image of the impacted foil emission recorded by the camera and a temporal profile of the emission is shown in Fig. 3. Based on the shock breakout time measurements the projectile velocity can be determined from the formula: $v_p \approx \beta L_{acc}/t_{acc}$ , where $L_{acc} = L_{Ch}$ , $t_{acc} = (t_{sb} - t_L) - (\Delta t_p + \Delta t_{sb})$, and $\Delta t_{sb}$ is the time needed to pass the shock through the impacted foil. For the case presented in Fig. 3, at $\beta \approx 1.36$ and $\Delta t_{sb} \approx 0.3$ ns (estimation based on numerical simulations) we obtain $v_p \approx 142$ km/s.

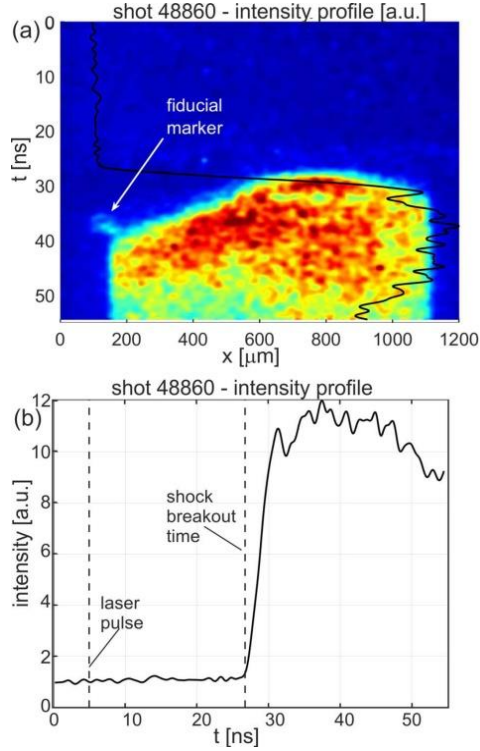

Fig. 3. The streak camera image (a) and the temporal intensity profile (b) of the light emission from the rear surface of the Al foil impacted by the gold plasma projectile driven by LICPA. The time delay between the fiducial and the laser pulse irradiating the CH/Au target is equal to 32.7 ns. $E_L$ = 232 J.

For the determination of the projectile parameters with the TOF method, a set-up of nine ion charge collectors (ICs) located in the horizontal plane approximately symmetrically relative to the accelerator axis within the angle of 39° and at the distance of 31 cm from the accelerator was used (Fig. 4). From the TOF spectra recorded by the collectors we were able to determine various characteristics of the projectile plasma at long distance from the accelerator, in particular: the angular distribution of plasma (ion) velocity averaged over time, the mean plasma velocity $v_p$ defined as the ion velocity averaged over time and angle, the angular distribution of ion charge density, the spatial distribution of ion charge density for different times and the total ion charge of plasma Q. In order to calculate $v_p$, we first calculated the ion charge density $q_\theta$ and the ion velocity averaged in time $v_\theta$ in the $\theta$ direction (using the TOF spectrum recorded in the $\theta$ direction), and then $v_p$ was calculated from the formula: $v_p = \sum^\theta v_\theta q_\theta / \sum^\theta q_\theta$. Having Q and $v_p$ and assuming that the projectile velocity is equal to the mean plasma velocity, the kinetic energy of the projectile can be calculated from the following formula: $E_p[J] \approx 10^{-8}(A/2Z)Qv_p^2$, where A is the mass number of plasma ions, Z is the average charge state of plasma ions and Q and $v_p$ are in the SI units (C, m/s). The value of Z at the long distance from the accelerator (where ICs were located) could not be calculated from the TOF measurements, and we were able to estimate only the value $Z_{out}$ at the accelerator channel outlet based on the 2D PALE simulations where the average charge state of plasma ions was calculated from the QEOS equations [29]. However, due to the free expansion of the plasma in space between the accelerator and the ICs, the plasma is adiabatically cooled and the plasma ions have enough time (>1ns) to recombine before reaching ICs. As a consequence, the average charge state of ions reaching ICs can be significantly lower than $Z_{out}$. Thus, having Q and $v_p$ from the TOF measurements and $Z_{out}$ from the simulations we can only determine a lower limit of the projectile kinetic energy and state that $E_p[J] > 10^{-8}(A/2Z_{out})Qv_p^2$. On the other hand, assuming the known mass of the accelerated plasma $m_p$, the kinetic energy can be determined from $E_p = (1/2)m_p v_p^2$, so in the same way as in case of the measurements using interferometry or the streak camera. However, in case of using the ion diagnostic, $v_p$ is determined directly from measurements without the necessity to determine the factor β characterizing the run of the acceleration process in the accelerator.

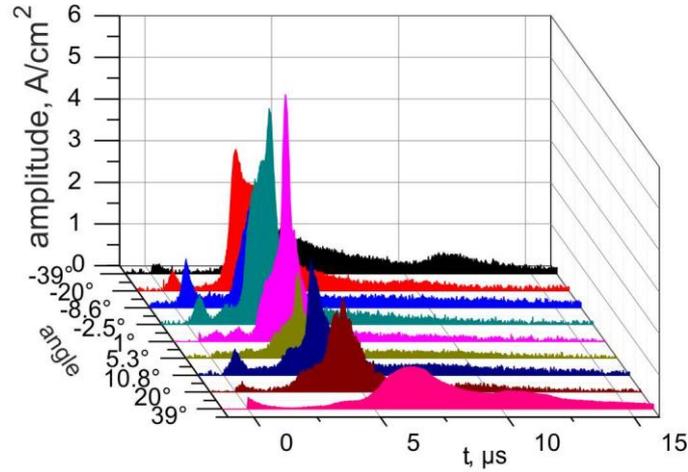

Fig. 4. The TOF spectra of the plasma projectile driven by LICPA recorded by the ion collectors at a long distance (31 cm) from the CH/Au target. $E_L$ = 200 J.

Fig. 4 presents sample TOF spectra of the projectile plasma recorded by ICs. In the presented case, almost all mass and kinetic energy of plasma is accumulated in the time period $t \leq 6$ us, and the plasma moves with the mean velocity $v_p \approx 135$ km/s.

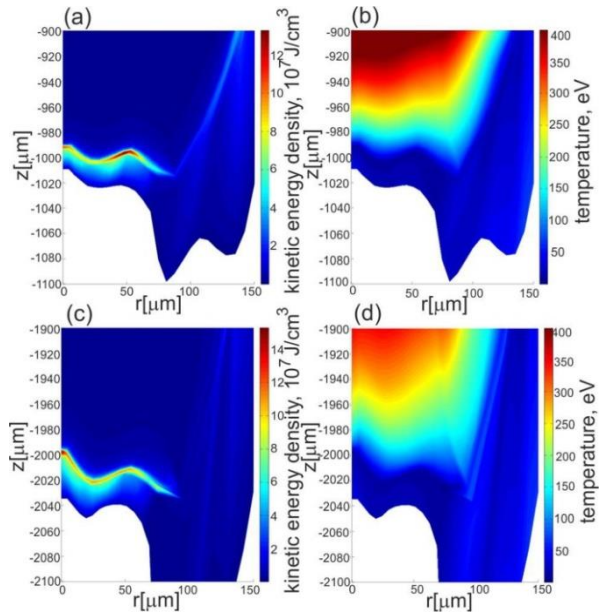

Fig. 5. 2D spatial profiles of kinetic energy density (a,c) and temperature (b,d) of plasma inside the LICPA accelerator channel at the moment when the gold projectile is in the middle of the channel length (z = -1000 μm) and when it begins to leave the channel (z = - 2000 μm). The CH/Au target placed at the beginning of the channel (z = 0) is irradiated by the laser beam introduced into the accelerator cavity and directed along the channel axis (r = 0). $E_L$ = 200 J.

Selected results of numerical simulations of the acceleration of the gold projectile in the LICPA scheme using the 2D PALE code are shown in Fig. 5. The figure presents 2D spatial distributions of the kinetic energy and temperature of plasma in the LICPA accelerator in two stages of the projectile acceleration: when the main part of the projectile reaches the distance z = -1000 μm (a, b) and when the projectile is close to the channel outlet (z = -2000 μm, figures c, d). It can be seen that for both stages the projectile is a compact plasma object of the peak density > 5 g/cm$^3$, and almost all (~ 90 %) kinetic energy and mass of the projectile is stored within a small z region of Δz < 100 μm (the plasma density distribution in this region, not shown here, is almost identical as the kinetic energy distribution). The temperature of gold plasma in the main, dense part of the projectile is relatively low and at the channel outlet its (averaged) value is about 50 eV and the average charge state of Au ions in the plasma (calculated from the OEOS equations) is approximately equal to $Z_{out} \approx 10$. However, due to the high projectile density, the pressure inside the projectile is high (~ 0.5 Mbar). As a result, when the projectile leaves the accelerator channel it not only moves along the z axis but also expands in radial direction, which is clearly visible in the TOF spectra (Fig. 4).

Table I summarizes the results of our measurements. It presents values of the mean velocity $v_p$ and the kinetic energy $E_p$ of the gold plasma projectile as well as the energetic acceleration efficiency $\eta = E_p/E_L$ inferred from the measurements using different diagnostics. For the estimation of kinetic energy of the projectile from the measurements using interferometry, the streak camera and the ion diagnostic in case of estimation 1, we assumed the plasma projectile mass equal to 0.9 of the gold disc mass $m_d$ = 4.1 μg (this assumption was based on a detailed analysis of results of simulation of the projectile acceleration using the PALE code). However, for the last case the projectile velocity $v_p$ was calculated from the ion diagnostic without any assumption on the coefficient β, while for the interferometry and the streak camera we assumed β = 1.36 ± 0.02 (see above). For the ion diagnostics, in case of estimation 2, the lower limit of the kinetic energy was calculated based on the formula $E_p[J] > 10^{-8}(A/2Z_{out})Qv_p^2$ with $Z_{out} \approx 10$ taken from the PALE simulations (see above). In this case any assumptions were made on the projectile mass or β coefficient (Q and $v_p$ were determined directly from the ion diagnostics).

Table I. The values of the mean velocity $v_p$ and the kinetic energy $E_p$ of the gold plasma projectile accelerated in the LICPA scheme or the AA scheme as well as the energetic acceleration efficiency $\eta = E_p/E_L$ inferred from the measurements using different diagnostics.

| | Projectile parameter | mean velocity [km/s] | kinetic energy [J] | acceleration efficiency [%] |
|---|---|---|---|---|
| LICPA | interferometry | 146±8 | 39±5 | 19.0±2.9 |
| | ion diagnostic estimation 1 | 135±15 | 34±8 | 16.9±3.8 |
| | ion diagnostic estimation 2 | 135±15 | > 29±6 | >14.4±3.3 |
| | streak camera | 137±7 | 35±4 | 15.1±1.8 |
| Ablative acceleration | interferometry | < 0.6 | < 7 | < 3 |
| | ion diagnostic | 35±10 | 2.3±1.3 | 1.2±0.7 |

The table demonstrates fairly high similarity of the parameters of the LICPA-driven projectile obtained with quite different measurement instruments and methods of the parameters estimation – within the experimental errors the average values of these parameters are approximately the same. In particular, the projectile is accelerated to the velocity 140 km/s with the energetic acceleration efficiency of 15 – 19 %. This acceleration efficiency is by more than factor 10 higher than the one obtained for the AA scheme (1.2 %) which, in turn, is comparable to those achieved in many other experiment performed so far [11-15,19].

Finally, we measured craters produced in the massive Al target by impact of the gold plasma projectile accelerated in the LICPA and AA schemes by the 3w, 200 J laser pulse. In both schemes the massive target was placed at the outlet of the cylindrical channel which in case of LICPA had the length of 2 mm and the diameter of 0.3 mm while for AA the channel diameter was equal to 2 mm (as previously) and the channel length was 0.4 mm, 1 mm or 2 mm. The volume of crater produced by the projectile driven by LICPA was equal to 8.5 mm$^3$, and was by a factor 25, 45 and 57 larger than the crater volumes obtained for the AA scheme with the channel length of 0.4 mm, 1 mm and 2 mm, respectively. Since the crater volume is proportional to the energy of the shock wave generated in the target by the projectile impact and the shock energy is the higher the higher is the projectile kinetic energy, the observed big difference in dimensions of craters produced with the use of LICPA and AA is a clear indirect evidence that the kinetic energy of the projectile driven by LICPA is much higher than those of the projectile accelerated by AA.

## 3. Conclusions

In conclusion, the acceleration of the heavy plasma projectile in the LICPA accelerator has been investigated. It has been shown that in the accelerator with a 200-J laser driver a 4-µg gold plasma projectile can be accelerated to the velocity of 140 km/s with the energetic acceleration efficiency of 15 – 19 % which is at least by a factor 5 -10 higher than those achieved in the ablative acceleration schemes and is the highest among the ones measured so far for any projectiles accelerated to the velocities ≥ 100 km/s. Such high acceleration efficiency coupled with the high density of the projectile driven by LICPA makes it possible to produce high-pressure shocks by the projectile impact much more efficiently than with the methods used so far [28]. As a result, to create high-energy-density matter states, e.g. for studies of the equation of state of various materials, much smaller and cheaper laser drivers could be used which would open the possibility to carry out such studies also in small laboratories.

**Acknowledgements**
This work was supported in part by the Ministry of Science and Higher Education, Poland under the grant no W39/7.PR/2015. Also, this work was supported in part by the Czech Ministry of Education, Youth and Sport , projects LD 14089 and RVO 68407700. The experiment was performed within the Access to Research Infrastructure activity in the 7$^{th}$ Framework Programme of EU (Contract No. 284464, Laserlab Europe-III, project pals002015).